\documentclass[12pt,reqno]{amsart}
\usepackage{amsmath,amssymb,amsfonts,amsthm}
\usepackage[mathscr]{eucal}
\usepackage[all]{xy}
\usepackage{hyperref}
\usepackage{setspace}
\allowdisplaybreaks

\begin{document}
\title{Geometrical model of massive spinning particle in four-dimensional Minkowski space}

\author{D. S. Kaparulin, S. L. Lyakhovich, I. A. Retuntsev} 
 \email{dsc@phys.tsu.ru, sll@phys.tsu.ru, retuntsev.i@phys.tsu.ru}

\address{ Physics Faculty, Tomsk State University, Tomsk 634050,
Russia }

\begin{abstract} We propose the model of massive spinning
particle traveling in four-dimensional Minkowski space. The
equations of motion of the particle follow from the fact that all
the classical paths of the particle lie on a cylinder whose position
in Minkowski space is determined by the particle's linear momentum
and total angular momentum. All the paths on one and the same
cylinder are gauge equivalent. The equations of motion are found in
implicit form for general time-like paths, and they are
non-Lagrangian. The explicit equations of motion are found for
trajectories with small curvature and helices. The momentum and
total angular momentum are expressed in terms of characteristics of
the path in all the cases. The constructed model of the spinning
particle has geometrical character, with no additional variables in
the space of spin states being introduced.
\end{abstract}

\maketitle

\section{Introduction}
The classical spinning particle is irreducible if its quantization
corresponds to the irreducible representation of the Poincare group.
The models of irreducible spinning particles are well-known
\cite{Fren, Univ, Bratek, Rempel}. In most of these models the
particle is considered as the point object in Minkowski space, while
the spin positions take values in the internal space. Depending on
the choice of the coordinates in internal space, vector, tensor,
spinor and twistor types of spinning particle models are
distinguished \cite{Fryd}. In the geometrical models the spin is
described in terms of derivatives of the particle's path, and no
internal space is introduced. The variational geometrical models are
known for massless particles \cite{Ramos1,Ramos2}. The variational
geometrical models of massive particles are reducible
\cite{Pisarski, Plyush}.

In the paper \cite{WS}, the concept of the world sheet is proposed
to describe massive particles with spin. The idea of this concept is
that the classical trajectories of irreducible particle lie on a
certain cylindrical surface in Minkowski space. The position of the
cylinder in space is determined by the values of the particle's
linear momentum and total angular momentum, and all the paths on one
and the same cylinder are gauge equivalent. The world sheet concept
can be used to deduce the ordinary differential equations describing
the particle's path. These equations are non-Lagrangian and involve
higher derivatives. The geometrical model of massive spinning
particle traveling light-like and isotropic paths in
three-dimensional Minkowski space was developed in the original work
\cite{WS}. In four-dimensional Minkowski space, the world sheet is
considered in \cite{RPJ}.

In the present research, we construct the geometrical model of
massive spinning particle traveling in four-dimensional Minkowski
space. The content of the article is based on the paper \cite{RPJ},
but more attention is paid to the geometry of particle's paths. In
particular, we consider the cylindrical curves with small curvature,
and also the helical paths. The construction of equations of
cylindrical curves uses several techniques of differential geometry,
including the Frenet moving frame. For differential geometry of the
curves in (non-)Euclidean space we refer to the books \cite{ModGeom,
Rosenfeld}. The method of classification of cylindrical curves in
Euclidean space has been proposed in \cite{Cylinder, Starostin}. We
use this method with small modifications accounting the signature of
the space-time metric.

The rest of the article is organized as follows. In Section 2, we
describe the geometry of the world sheet of massive spinning
particle in four-dimensional Minkowski space. In Section 3, we
derive the equations for general cylindrical path on the world sheet
with time-like tangent vector. The equations are derived in implicit
form. In Section 4, some particular types of cylindrical curves with
explicit equations of motion are considered. The conclusion
summarizes the results.

\section{World sheets and world paths}
The geometry of the world sheet of massive spinning particle in
four-dimensional space-time has been described in ref. \cite{RPJ}.
It has been shown that the world surface is determined by two
algebraic equations
\begin{equation}\label{Cyl-eq}
    \phantom{\frac12}(x-y)^2+(n,x)^2=R^2\,,\qquad (x-y,a)=\Delta\,,\phantom{\frac12}
\end{equation}
where $x^\mu\,,\mu=\overline{0,3}$ are the coordinates in Minkowski
space, and $R,\Delta$ are constants. The vector parameters
$n,\,y,\,a$, are subjected to the constraints
\begin{equation}\label{nya-constr}
    \phantom{\frac12}(n,n)=-1\,,\qquad (a,a)=1\,,\qquad (n,y)=(a,y)=0\,.\phantom{\frac12}
\end{equation}
Hereinafter, the round brackets denote the scalar product in
Minkowski space,
\begin{equation}\label{}
    \phantom{\frac12}(u,w)=u_\mu w^\mu\,,\qquad \forall u,w\,.\phantom{\frac12}
\end{equation}
We use the mostly positive convention for the signature of the
metric throughout the paper. All the world indices are raised and
lowered by the metric.

The geometric meaning of the equations (\ref{Cyl-eq}) is following.
The first equation determine the circular hypercylinder with
time-like axis in Minkowski space. The tangent vector to the
cylinder axis is $n$, and the vector y connects the cylinder axis
and the origin by the shortest path. The radius of hypercilder is
$R$. The second of equations (\ref{Cyl-eq}) determines the
hyperplane with the normal. The hypercylinder axis is tangent to
this hyperplane. The distance between hypercylinder axis and
hyperplane is $\Delta$. The intersection of the hypercylinder and
hyperplane is two-dimensional cylinder. The tangent vector to the
cylinder axis is $n$. The radius of the cylinder is
\begin{equation}\label{}
    \phantom{\frac12}r=\sqrt{R^2-\Delta^2}\,.\phantom{\frac12}
\end{equation}
It is a real quantity if
\begin{equation}\label{}
    \phantom{\frac12}0\leq|\Delta|\leq R\,.\phantom{\frac12}
\end{equation}
Otherwise, the intersection of the hypercylinder and hyperplane is
empty. In the special case $r=0$, the cylinder becomes a straight
line. We dot not consider this possibility below.

The linear momentum $p=p_\mu dx^\mu$ and total angular momentum
$J=J_{\mu\nu} dx^\mu \wedge dx^\mu$ of the particle are determined
by the world sheet position in the Minkowski space,
\begin{equation}\label{pJ-nya}
    \phantom{\frac12}p=mn\,,\qquad J=m y\wedge
    n+s\varepsilon_{\mu\nu\rho\sigma}n^\mu a^\nu dx^\rho\wedge
    dx^\sigma\,,\phantom{\frac12}
\end{equation}
where $\wedge$ denotes the exterior product, and
$\varepsilon_{\mu\nu\rho\sigma},\varepsilon_{0123}=1$ is the
Levi-Civita symbol. The constants $m,s$ are the mass and spin of the
particle. The details of the derivation of the formulas
(\ref{pJ-nya}) are given in \cite{RPJ}. With account of relations
(\ref{nya-constr}), the states of the spinning particle belong to
the co-orbit of the Poincare group,
\begin{equation}\label{co-orbit}
    \phantom{\frac12}p^2+m^2=0\,,\qquad w^2=m^2s^2\,,\phantom{\frac12}
\end{equation}
where $w=sa$ is the Pauli-Lyubanski pseudovector. The spin tensor is
determined by the standard rule $S=J-x \wedge p$, form which we find
\begin{equation}\label{}
    S=m (y-x) \wedge n+s\varepsilon_{\mu\nu\rho\sigma}n^\mu a^\nu dx^\rho\wedge
    dx^\sigma\,.
\end{equation}
By construction,
\begin{equation}\label{}
    \frac{1}{2}S_{\mu\nu}S^{\mu\nu}=s^2-m^2R^2\,,\qquad
    \frac{1}{8}\varepsilon_{\mu\nu\rho\sigma}S^{\mu\nu}S^{\rho\sigma}=-ms\Delta\,.
\end{equation}
The right hand sides of these equalities are constants because of
irreducibility conditions of the Poincare group representation
\cite{RPJ}.

The classical trajectories of spinning particles are curves on the
world sheet. The casuality condition $\dot{x}^0>0$ is imposed on the
physical path. We term the path locally regular if each its small
segment lies on a unique representative in set of cylinders. Almost
all the cylindrical paths are locally regular. Only locally regular
paths are considered in this article.

\section{Classification of cylindrical curves}
In this section, we obtain the system of equations describing the
cylindrical curves in four-dimensional Minkowski space.

The starting point of our consideration is that each cylindrical
curve $x=x(\tau)$ lies on a certain cylinder in the set
(\ref{Cyl-eq}). In this case, the equations (\ref{Cyl-eq}) are
identically satisfied for all the values of the parameter $\tau$ on
the curve. The differential consequences of these identities have
the form
\begin{equation}\label{Cyl-dc}
    \frac{d^k}{d\tau^k}\bigg((x-y)^2+(n,x)^2-R^2\bigg)=0\,,\qquad
    \frac{d^k}{d\tau^k}\bigg((x-y,a)-\Delta\bigg)=0\,,\qquad
    k=1,2,3,\ldots\,.
\end{equation}
Once the differential consequences of sufficiently large order are
included, the cylinder parameters can be expressed as functions of
derivatives of the path,
\begin{equation}\label{nya-xdx}
    \phantom{\frac12}n=n(x,\dot{x},\ddot{x},\ldots)\,,\qquad
    y=y(x,\dot{x},\ddot{x},\ldots)\,,\qquad
    a=a(x,\dot{x},\ddot{x},\ldots)\,,\phantom{\frac12}
\end{equation}
where the dots denote the derivatives by $\tau$. The existence of
unique solution (\ref{nya-xdx}) of the equations (\ref{Cyl-dc}) uses
the regularity assumption for the cylinder path. For singular path,
the solution is not necessary unique. That is why the singular
cylindrical paths are excluded. The substitution of the cylinder
parameters (\ref{nya-xdx}) into the original equations gives us the
equations for the cylindrical curves. The cylinder parameters are
integrals of motion of this equation, which can be treated as the
equation of motion of spinning particle. The linear momentum and
angular momentum of the particle are defined by the rule
(\ref{pJ-nya}).

We can proceed with explicit derivation of the equations of
cylindrical path. It is sufficient to differentiate equations
(\ref{Cyl-eq}) four times. In this case, the differential
consequences (\ref{Cyl-dc}) have the form
\begin{equation}\label{hypp-dc}\begin{array}{c}\displaystyle
    (\dot{x},a)=0\,,\qquad (\ddot{x},a)=0\,,\qquad
    (\dddot{x},a)=0\,,\qquad (\ddddot{x},a)=0\,;
\end{array}\end{equation}

\begin{equation}\label{hypc-dc}\begin{array}{c}\displaystyle
    (\dot{x},d)=0\,,\qquad (\ddot{x},d)+(\dot{x},n)^2=0\,,\qquad
    (\dddot{x},d)+3(\ddot{x},n)(\dot{x},n)=0\,,\\[6mm]\displaystyle
    (\ddddot{x},d)+4(\dddot{x},n)(\dot{x},n)+3(\ddot{x},n)^2=0\,,
\end{array}\end{equation}
where the notation is used,
\begin{equation}\label{d-def}
    d\equiv x+n(n,x)-y\,.
\end{equation}
The vector $d$ connects the current position of the particle and
hypercylinder axis by the shortest path. To simplify the formulas,
we assume that the path is time-like and the parameter $\tau$ is
natural on the curve,
\begin{equation}\label{}
    (\dot{x},\dot{x})=-1\,.
\end{equation}
The relations (\ref{hypp-dc}), (\ref{hypc-dc}) constitute the system
of equations (\ref{Cyl-dc}) for finding of cylinder parameters
$n,y,a$.

The normalized vector $a$ can be expressed from equations
(\ref{hypp-dc}),
\begin{equation}\label{}
    a=\pm\frac{[\dot{x},\ddot{x},\dddot{x}]}{\sqrt{[\dot{x},\ddot{x},\dddot{x}]^2}}\,,\qquad
    [\dot{x},\ddot{x},\dddot{x}]\equiv\varepsilon_{\mu\nu\rho\sigma}\dot{x}^\mu\ddot{x}^\mu\dddot{x}^\rho
    dx^\sigma\,,
\end{equation}
where the square brackets denote the vector product of three vectors
in Minkowski space. The consistency condition for equations
(\ref{hypp-dc}) reads
\begin{equation}\label{zero-curv}
    ([\dot{x},\ddot{x},\dddot{x}]\,,\ddddot{x})\equiv(\dot{x},\ddot{x},\dddot{x},\ddddot{x})=0\,,
\end{equation}
where the round brackets denote the mixed product of four vectors in
Minkowski space. From the geometrical viewpoint, the obtained
condition implies zero torsion of the curve $x(\tau)$. It ensures
that this curve lies in some hyperplane, with $a$ being its normal.

Now we consider the hypercylinder condition and its differential
consequences. We have nine unused conditions (\ref{Cyl-eq}),
(\ref{nya-constr}), (\ref{hypc-dc}) to express two vectors $n,y$ and
find a single consistency condition. Introduce the Frenet moving
frame for time-like curves \cite{Rosenfeld},
\begin{equation}\label{Frenet-Frame}\begin{array}{c}\displaystyle
    e_1=\dot{x}\,,\qquad
    e_2=\frac{\ddot{x}}{\sqrt{(\ddot{x},\ddot{x})}}\,,\\[7mm]\displaystyle
    e_3=\frac{(\ddot{x},\ddot{x})^2\dot{x}+(\ddot{x},\dddot{x})\ddot{x}-(\ddot{x},\ddot{x})\dddot{x}}
    {\sqrt{(\dddot{x},\dddot{x})(\ddot{x},\ddot{x})^2+(\ddot{x},\ddot{x})^4-(\ddot{x},\dddot{x})^2(\ddot{x},\ddot{x})}}\,,\qquad
    e_4=\frac{[\dot{x},\ddot{x},\dddot{x}]}{\sqrt{[\dot{x},\ddot{x},\dddot{x}]^2}}\,.
\end{array}\end{equation} All the basis vectors of Frenet frame are
normalized and orthogonal to each other,
\begin{equation}\label{Frenet-Frame-sp}\begin{array}{c}\displaystyle
    \phantom{\frac12}(e_1,e_1)=-1\,,\qquad
    (e_2,e_2)=(e_3,e_3)=(e_4,e_4)=1\,,\phantom{\frac12}\\[3mm]\displaystyle
    \phantom{\frac12}(e_i,e_j)=0\,,\qquad i\neq j\,.\phantom{\frac12}
\end{array}\end{equation}
The curvatures $k_1,k_2$, and torsion $k_3$ of the curve are
determined by the rule
\begin{equation}\label{k1k2k3}\begin{array}{c}\displaystyle
    \phantom{\frac12}k_1=\sqrt{(\ddot{x},\ddot{x})}\,,\qquad
    k_2=-\frac{1}{(\ddot{x},\ddot{x})}\sqrt{(\dddot{x},\dddot{x})(\ddot{x},\ddot{x})
    -(\ddot{x},\dddot{x})^2-(\ddot{x},\ddot{x})^3}\,,\phantom{\frac12}\\[5mm]\displaystyle
    \phantom{\frac12}k_3=\frac{(\dot{x},\ddot{x},\dddot{x},\dddot{x})}{\sqrt{((\dddot{x},\dddot{x})
    -(\ddot{x},\dddot{x})^2(\ddot{x},\ddot{x})^{-1}-(\ddot{x},\ddot{x})^2)[\dot{x},\ddot{x},\dddot{x}]^2}}\,.\phantom{\frac12}
\end{array}\end{equation} The time derivatives of particle's position can be
expressed as the linear combinations of the Frenet basis vectors
(\ref{Frenet-Frame}) with the coefficients depending on the
curvatures and torsion of path,
\begin{equation}\label{dx-e}\begin{array}{c}\displaystyle
    \phantom{\frac12}\ddot{x}=e_1\,,\qquad \ddot{x}=k_1e_2\,,\qquad
    \dddot{x}=k_1^2 e_1+\dot{k}_1e_2+k_1k_2e_3\,,\phantom{\frac12}\\[5mm]\displaystyle
    \phantom{\frac12}\ddddot{x}=3\dot{k}_1k_1e_1+(\ddot{k}_1+k_1^3-k_1
    k_2^2)e_2+(2\dot{k}_1k_2+k_1\dot{k}_2)e_3+k_1k_2k_3e_4\,.\phantom{\frac12}
\end{array}\end{equation}
The unknown vectors $n,d$ admit the following represenation:
\begin{equation}\label{nd-e}
    \phantom{\frac12} n=\beta_1e_1+\beta_2(-\alpha_2e_2+\alpha_1e_3)\,,\qquad
    d=r(\alpha_1e_2+\alpha_2e_3)\pm\Delta e_4\,,\phantom{\frac12}
\end{equation}
where $\alpha_1,\alpha_2,\beta_1,\beta_2$ are new unknowns.
Substitution of the expressions (\ref{dx-e}), (\ref{nd-e}) into
(\ref{Cyl-eq}), (\ref{nya-constr}), (\ref{hypc-dc}) brings us the
following system of algebraic equations:
\begin{equation}\label{aabb}\begin{array}{c}\displaystyle
    \alpha_1^2+\alpha_2^2=1\,,\qquad \beta_1^2-\beta_2^2=1\,,\qquad
    r k_1 \alpha_1+\beta_2^2=0\,,\qquad
    r(\dot{k}_1\alpha_1+k_1k_2\alpha_2)+3k_1\alpha_2\beta_1\beta_2=0\,,\\[5mm]\displaystyle
    r\bigg[(\ddot{k}_1+k_1^3-k_1k_2^2)\alpha_1+(2\dot{k}_1k_2+k_1\dot{k}_2)\alpha_2\bigg]+
    4(\dot{k}_1\alpha_2-k_1k_2\alpha_1)\beta_1\beta_2-k_1^2((3\alpha_1^2-7)\beta_2^2-3)=0\,.
\end{array}\end{equation}
The other conditions are automatically satisfied.

The general method of solution of systems of algebraic equations
uses techniques of resultants. In this approach, all the variables
except one are eliminated, while all the other unknowns are
expressed as functions of this single variable from subresultants.
For system (\ref{aabb}), $\beta_1$ is a good candidate for the
independent variable. It has the sense of the hyperbolic tangent of
the angle $\theta$ between the element of cylinder and particle's
path,
\begin{equation}\label{}
    \beta_1=\text{ch}\,\theta\,.
\end{equation}
Eliminating the variables $\alpha_1,\alpha_2,\beta_2$ from the
system (\ref{aabb}) in two different ways, we get two algebraic
equations of order $16$ and $24$ for $\beta_1$,
\begin{equation}\label{P1P2}\begin{array}{rl}\displaystyle
P_1(\beta_1)&=81k_{1}^{4}\beta_{1}^{16}-162k_{1}^{4}\beta_{1}^{14}+r^2(18k_{1}^{4}k_{2}^2-18\dot{k}{}_{1}^{2}k_{1}^2+81k_{1}^4)\beta_{1}^{12}+r^2(18k_{1}^{2}k_{1}^{2}-18k_{1}^{4}k_{2}^2)\beta_{1}^{10}+\\[5mm]\displaystyle
&\hspace{-13mm}+r^4(k_{1}^{4}k_{2}^{4}+\dot{k}{}_{1}^{4}+2\dot{k}{}_{1}^{2}k_{1}^{2}k_{2}^{2}-18k_{1}^{6}k_{2}^{2})\beta_{1}^{8}
+r^{4}18k_{1}^{6}k_{2}^{2}\beta_{1}^{6}-r^6(2\dot{k}{}_{1}^{2}k_{1}^{4}k_{2}^{2}+2k_{1}^{6}k_{2}^{4})\beta_{1}^{4}+r^{8}k_{1}^{8}k_{2}^{4}=0\,,\\[5mm]\displaystyle
P_2(\beta_1)&=81k_{1}^{4}\beta_{1}^{24}+324k_{1}^{4}\beta_{1}^{22}+\textbf{260
    terms}=0\,.
\end{array}\end{equation}
The unknown $\beta_1$ is the positive common root of these
polynomials. We can't express it in the explicit form, but we can
argue that it is a function of first and second curvatures $k_1,\,
k_2$ with its derivatives,
\begin{equation}\label{}
    \beta_1=\beta_1(k_1,k_2,\dot{k}_1,\dot{k}_2,\ddot{k}_1)\,.
\end{equation}
As the curvatures are functions of the derivatives of the path, this
solution involves derivatives of $x$ up to fourth order. The other
unknowns are expressed using $\beta_1$,
\begin{equation}\label{aab}\begin{array}{c}\displaystyle
    \alpha_1=\frac{1-\beta_1^2}{rk_1}\,,\qquad
    \alpha_2=\frac{2r\dot{k}_1k_1^{-2}k_2(1-\beta_1^2)((1-\beta_1^2)^2-r^2k_1^2)}
    {r^2k_1^2(1-\beta_1^2)^2+k_1^2(9\beta_1^2(1-\beta_1^2)-r^2
    k_2^2)((1-\beta_1^2)^2-r^2k_1^2)}\,,\\[7mm]\displaystyle
    \beta_2=-\frac{r^2k_1^2(1-\beta_1^2)^2+k_1^2(9\beta_1^2(1-\beta_1^2)-r^2
    k_2^2)((1-\beta_1^2)^2-r^2k_1^2)}{6rk_1^2k_2\beta_1((1-\beta_1^2)^2-r^2k_1^2)}\,.
\end{array}\end{equation}
The momentum and total angular momentum of the particle are
expressed from equations (\ref{pJ-nya}) and (\ref{aab}),
\begin{equation}\label{pJ}\begin{array}{rl}
    p&\displaystyle=\beta_1e_1+\frac{\dot{k}_1(1-\beta_1^2)}{3rk_1^4\beta_1}e_2-\\[5mm]&-\displaystyle
    \frac{r^2k_1^2(1-\beta_1^2)^3+k_1^2(1-\beta_1^2)(9\beta_1^2(1-\beta_1^2)-r^2
    k_2^2)((1-\beta_1^2)^2-r^2k_1^2)}{6r^2k_1^3k_2\beta_1((1-\beta_1^2)^2-r^2k_1^2)}e_3\,,
    \\[5mm]\displaystyle
    J&\displaystyle =m (x-d)\wedge n\pm s\bigg[\beta_1 e_2\wedge e_3-\frac{\dot{k}_1(1-\beta_1^2)}{3rk_1^4\beta_1}e_1\wedge e_3-\\[5mm]&-\displaystyle
    \frac{r^2k_1^2(1-\beta_1^2)^3+k_1^2(1-\beta_1^2)(9\beta_1^2(1-\beta_1^2)-r^2
    k_2^2)((1-\beta_1^2)^2-r^2k_1^2)}{6r^2k_1^3k_2\beta_1((1-\beta_1^2)^2-r^2k_1^2)}e_1\wedge
    e_2\bigg]\,.
\end{array}\end{equation}
By construction, $p$ and $J$ belong to the co-orbit (\ref{co-orbit})
of the Poincare group. This ensures the irreducibility of the
particle that moves the cylindrical path. The equation of
hypercylindrical curves is the consistency condition of equations
(\ref{P1P2}). It is given by the resultant of two polynomials,
\begin{equation}\label{Res}
    \text{Res}_{\beta_1}\,(P_1,P_2)=0\,.
\end{equation}
The explicit expression of this resultant is too long and not
informative. As a matter of principle, we mention that the resultant
involves the derivatives of the path up to fourth order. So the
curves on the hypercylinder are described by the fourth-order
equation.

The paths of spinning particle lie on the intersection of
hypercylinder and hyperplane. Such paths meet both the conditions
(\ref{zero-curv}) and (\ref{Res}). This is a system of two equations
of fourth order. The model of spinning particle, being described by
these equations, has the geometrical character because no additional
variables in the spin sector are introduced.

\section{Some particular cases}
Let us consider some particular cases of cylindrical curves such
that the system (\ref{aabb}) can be solved explicitly.

At first, suppose that the path of particle is close to the straight
line. We solve equations (\ref{aabb}) using the following
assumptions. The curvatures and derivatives are small quantities of
same order of magnitude:
\begin{equation}\label{rk}
    r\,k_1\sim r\, k_2\sim r^2\,\dot{k}_1\sim r^2\dot{k}_2\sim r^3\,
    \ddot{k}_1\ll 1\,.
\end{equation}
The parameters $\alpha_1,\,\alpha_2,\,\beta_1,\,\beta_2$ are sought
in the form
\begin{equation}\label{aabb-ansatz}\begin{array}{c}\displaystyle
    \alpha_2=-1+o(\varphi)\,,\qquad
    \alpha_1=\varphi+o(\varphi^2)\,,\qquad
    \beta_1=1+o(\theta)\,,\qquad
    \beta_2=\theta+o(\theta^2)\,,
\end{array}\end{equation}
where the parameters $\varphi,\theta$ have the same order of
magnitude as (\ref{rk}).

Substituting the ansatz (\ref{aabb-ansatz}) into the equations
(\ref{aabb}) and accounting (\ref{rk}), we get the following
equations:
\begin{equation}\label{aabb-approx}
    rk_1\varphi+\theta^2=0\,,\qquad
    r\dot{k}_1\varphi+3k_1\theta=rk_1k_2\,,\qquad
    r\ddot{k}_1\varphi-4\dot{k}_1\theta=(2\dot{k}_1k_2+k_1\dot{k}_2)-3k_1^2\,.
\end{equation}
This system includes three equations for two unknowns
$\varphi,\theta$. These quantities can be expressed from the linear
equations,
\begin{equation}\label{}
    \varphi=\frac{3r^{-1}k_1^3+k_1(10\dot{k}_1k_2+3k_1\dot{k}_2)}{3\ddot{k}_1k_1+4\dot{k}_1^2},\qquad
    \theta=\frac{3r^{-1}k_1^2\dot{k}_1-k_1 (\dot{k}_1 k_2+k_1
    \dot{k}_2)}{3\ddot{k}_1k_1+4\dot{k}_1^2}\,.
\end{equation}
This solution exists if the regularity condition for the curve is
satisfied,
\begin{equation}\label{}
\phantom{\frac12}3\ddot{k}_1k_1+4\dot{k}_1^2\neq0\,.\phantom{\frac12}
\end{equation}
The consistency condition for the equations of motion
(\ref{aabb-approx}) has the form
\begin{equation}\label{Res-app}
    \phantom{\frac12}3k_1^4+rk_1^2(10\dot{k}_1k_2+3k_1\dot{k}_2)(3\ddot{k}_1k_1+4\dot{k}_1^2)+(3r^{-1}k_1^2\dot{k}_1-k_1 (\dot{k}_1 k_2+k_1
    \dot{k}_2))^2=0\,.\phantom{\frac12}
\end{equation}
This forth order equation for the hypercylindrical curves with small
curvature.

The momentum and angular momentum of the particle moving the path
with small curvature have the form
\begin{equation}\label{}
    p=me_1+o(1)\,,\qquad J=m (x+re_2)\wedge e_1\pm se_2\wedge
    e_3+o(1)\,,
\end{equation}
where the terms labeled $o(1)$ vanish in the straight line
approximation. The obtained formulas are analogue of the formulas
(\ref{pJ}) in the small curvature limit. The system of equations
(\ref{zero-curv}), (\ref{Res-app}) is a system of two equations of
motion of the massive spinning particle. We can explicitly see that
both equations have the fourth order in derivatives of the path.

The other type of cylindrical curves are helices. The helices on
$2d$ cylinder are defined by the constant curvature conditions
\begin{equation}\label{}
    k_1,k_2=\text{const}\,.
\end{equation}
In this case, the system of equations (\ref{aabb}) takes the form
\begin{equation}\label{aabb-helix}\begin{array}{c}
     \alpha_{1}^2+\alpha_{2}^2=1\,,\qquad
     \beta_{1}^2-\beta_{2}^2=1,\qquad rk_{1}\alpha_{1}+\beta_{2}^2=0,\qquad
     rk_{1}k_{2}\alpha_{2}+3k_{1}\alpha_{2}\beta_{1}\beta_{2}=0,\\[5mm]\displaystyle
     r(k_{1}^3-k_{1}k_{2}^2)\alpha_{1}-4k_{1}k_{2}\alpha_{1}\beta_{1}\beta_{2}-k_{1}^2(3\alpha_{1}^2\beta_{2}^2-7\beta_{2}^2-3)=0.
\end{array}\end{equation}
The solution to this system reads
\begin{equation}\label{aabb-helix-sol}
    \alpha_1=-1,\qquad \alpha_2=0,\qquad
    \beta_1=\frac{k_2}{\sqrt{k_2^2-k_1^2}},\qquad
    \beta_2=\frac{-k_1}{\sqrt{k_2^2-k_1^2}}\,.
\end{equation}
The consistency condition for this system is the relation between
the curvatures of the helix and its radius,
\begin{equation}\label{}
    r=\frac{k_1}{k_2^2-k_1^2}\,.
\end{equation}
In this way, we see that the helices are particular solutions to the
equations of cylindrical curves (\ref{zero-curv}), (\ref{Res}).

The linear momentum and total angular momentum of the spinning
particle traveling the helical path have the form
\begin{equation}\label{}\begin{array}{rl}
    p&\displaystyle=m\bigg(\frac{k_2}{\sqrt{k_2^2-k_1^2}}e_1-\frac{k_1}{\sqrt{k_2^2-k_1^2}}e_3\bigg)\,,\\[5mm]
    J&\displaystyle=m(x+re_2)\wedge\bigg(\frac{k_2}{\sqrt{k_2^2-k_1^2}}e_1-\frac{k_1}{\sqrt{k_2^2-k_1^2}}e_3\bigg)\pm
    s\bigg(\frac{k_2}{\sqrt{k_2^2-k_1^2}}e_2-\frac{k_1}{\sqrt{k_2^2-k_1^2}}e_1\bigg)\wedge
    e_3\,.
\end{array}\end{equation}
By construction, the momentum and total angular momentum of the
particle lie on the co-orbit of the Poincare group.

\section{Conclusion}
In this article, we considered a geometrical model of irreducible
massive spinning particle. Our initial assumption was that the
classical paths of spinning particle lie on a certain cylinder whose
position is determined by the values of the particle's linear
momentum and total angular momentum. Preceding from the fact that
all the paths on one and the same cylinder are gauge equivalent, we
derived the particle's  equations of motion. The equations of motion
are found only in implicit form. The structure of these equations
brings us to the conclusion that the path of spinning particle
subjects to the system of two equations of forth order. The explicit
structure of these equations was studied in two particular cases of
the paths with small curvatures and helical trajectories. The
null-like paths are null-like helices, and they are possible path of
irreducible spinning particle. The model of massive spinning
particle traveling isotropic helical path was proposed earlier from
other considerations \cite{Ners}. The space-like cylindrical
trajectories can be studied in the similar way. We plan to study
this problem later.

\section*{Acknowledgements}
This research was funded by state task of Ministry of Science and
Higher Education of Russian Federation, grant number
3.9594.2017/8.9.

\end{document}